\documentclass[conference,a4paper]{IEEEtran}

\usepackage[utf8]{inputenc} 
\usepackage[T1]{fontenc}
\usepackage{url}
\usepackage{ifthen}
\usepackage{cite}
\usepackage[cmex10]{amsmath} 

\usepackage{color}
\usepackage{mathtools,amssymb,comment}

\newtheorem{lemma}{Lemma}
\newtheorem{theorem}{Theorem}

\newtheorem{claim}{Claim}

\newtheorem{example}{Example}
\newtheorem{remark}{Remark}

\begin{document}
\title{How to Use Undiscovered Information Inequalities:  Direct Applications of the Copy Lemma} 


\author{%
  \IEEEauthorblockN{Emirhan G\"urp{\i}nar}
  \IEEEauthorblockA{%
  		   ENS de Lyon, France\\
                    Email: emirhan.gurpinar@ens-lyon.fr}
  \and
  \IEEEauthorblockN{Andrei Romashchenko}
  \IEEEauthorblockA{%
  		  LIRMM, University of Montpellier, CNRS, France\\
		  on leave from IITP RAS\\
                    Email: andrei.romashchenko@lirmm.fr}
}

\maketitle

\begin{abstract}
We discuss linear programming techniques that help to deduce  corollaries of non-classical inequalities for Shannon's entropy.
We focus on direct applications of the copy lemma. 
These applications involve  implicitly some (known or unknown) non-classical  universal
inequalities for Shannon's  entropy,  though we do not  derive these  inequalities explicitly. 
To reduce the computational complexity of these problems, we extensively use symmetry considerations.

We present two examples of use of these techniques: we provide a reduced size formal inference of the best known bound for the Ingleton score
(originally proven by Dougherty \emph{et al.} with explicitly derived non-Shannon type inequalities), 
and improve the lower bound for the optimal information ratio of the secret sharing scheme for an access structure  based on the V\'amos matroid.
\end{abstract}

\section{Introduction}

We can associate with an $n$-tuple of jointly distributed random variables its \emph{entropy profile} that consists of 
the $2^n-1$ values of Shannon's entropy of each sub-tuple of the given tuple.
We say that a point in $\mathbb{R}^{2^n-1}$ is \emph{entropic} if it represents the entropy profile of some distribution.
The entropic points satisfy different information inequalities --- the constraints that characterize
the range of admissible entropies for jointly  distributed variables. The most known and well understood information inequalities
are so-called \emph{Shannon type inequalities} that are defined as linear combinations of several instances of the
\emph{basic inequality} $I(A:B|C)\ge 0$, where $A,B,C$ are any subsets (possible empty) of random variables.

In 1998 Z.~Zhang and R.W.~Yeung  discovered the first example of an (unconditional) non-Shannon type information inequality, 
which was a linear inequality for entropies of a quadruple of random variables that cannot be represented as a combination
of basic inequalities \cite{zy98}.  Now many other  non-Shannon type information are known (including several infinite
families of non-Shannon type information inequality), see, e.g., \cite{inf-ineq,ak-lemma-ineq,dougherty}.

The works on non-Shannon type information inequalities are focused on the following fundamental question: can we prove some
particular property for Shannon's entropy (usually expressed as an equality or an inequality),  assuming some specific constraints (which are also
given as equalities and inequalities for Shannon's entropy).
Not surprisingly, these arguments  typically use the technique of linear programming:  at the first stage,  to derive new information inequalities
(and to prove that they are indeed new) and then, at the second stage, to apply the derived inequalities  to some specific problem 
(e.g., in secret sharing,  network coding, and so on). Many applications of this type involve heavy computations: the number of 
information inequalities in $n$ random variables grows exponentially with $n$. 
These computations are typically performed with help of a computer, see, e.g.,  \cite{metcalf-burton,dougherty,padro-et-al-2010,tarik2018}.
It seems that in several problems of information theory the progress has slowed down since
the computational complexity of the corresponding linear programs is too high for the modern computers. So the question arises: 
can we reduce the computational complexity of the  relevant problems of linear programming? In this paper we show that in some
cases progress can be made with a reasonable combination of previously known techniques.

Recently Farr\`as  \emph{et al.}, \cite{tarik2018}, observed that  the two stages of the scheme mentioned above 
(deriving new information inequalities and applying them) 
can be merged in a unique problem of linear programming: in a specific application
of information inequalities, instead of enumerating explicitly all known non-Shannon information inequalities, we can  include in a linear program 
the \emph{constraints} from which those inequalities are (or can be) derived. The \emph{constraints} used in  \cite{tarik2018}
were instances of the Ahlswede--K\"orner lemma, \cite{ak-lemma}. As a tool of inference of non-Shannon type inequalities,  the Ahlswede--K\"orner lemma
is essentially equivalent to the Copy Lemma from \cite{dougherty} ``cloning'' one single random variable, 
see \cite{tarik-equiv}. In this paper we use a technique similar to \cite{tarik2018}, but  employing the \emph{constraints} obtained directly from
the Copy Lemma. This permits us to extend the technique: we apply the Copy Lemma to make  ``clones''  of  pairs of jointly distributed random variables
(instead of copies of individual random variables used in \cite{tarik2018}). 

To reduce the dimension of the corresponding problem of linear programming, we use symmetries of the problems  under consideration. 
Our usage of symmetries is similar to the ideas proposed in \cite{apte-walsh}, where the authors suggested to use symmetries   to  reduce  the
complexity of computing the outer bounds on network coding capacity. For a study of the general problem of Shannon and non-Shannon type inequalities for symmetric entropy  points we refer the reader to \cite{qi-yeung}.

To illustrate the power of this technique,  we discuss two examples. We start with a previously known result:  in the first example
we discuss a formal inference of the best known bound for the  Ingleton score (originally proven in \cite{dougherty}). 
We show that this bound can be achieved with only four auxiliary random variable and three applications of the Copy Lemma.
The obtained computer-assisted proof is very fast, and the ``formally checkable''
version of the proof can be reduced to a combination of 167 equalities and inequalities with rational coefficients. 
In our second example (the main result of the paper) we 
improve the known lower bound for the optimal information ratio of the secret sharing schemes associated with the V\'amos matroid 
(for the \emph{access structure} denoted $V_0$).

Though the techniques used in this paper are rather simple, their combination proves to be fruitful. We believe that this approach can be used in other problems of information theory.

\subsection*{Notation:}
We denote $[n]=\{1,\ldots,n\}$. Given a tuple of jointly distributed random variable $(X_1,\ldots, X_n)$
and a set of indices $V=\{i_1,\ldots,l_k\}\subset [n]$ we denote by $H(X_V)$ the Shannon entropy 
$H(X_{i_1},\ldots,X_{i_k})$.
We fix an arbitrary order on non-empty subsets of the set of indices $[n]$ and assign to each distribution 
$\mathbf{X}=(X_1,\ldots, X_n)$ its \emph{entropy profile}, i.e., 
the vector of $2^n-1$ entropies $H(X_V)$ (for all non-empty $V\subset [n]$).

We say that a point in the space $V_n=\mathbb{R}^{2^n-1}$ is \emph{entropic} if it represents the entropy profile
of some distribution. Following \cite{zy98}, we denote by $\Gamma_n^*$ the set of all entropic points.
We denote by $\bar \Gamma_n^*$ the topological close $\Gamma_n^*$; its elements are called \emph{almost entropic}.

We use the standard abbreviations for linear combinations of coordinates in the entropy profile:
 \[
 \begin{array}{lcl}
  H(X_V|X_W) &:=& H(X_{V\cup W}) - H(X_W) \\
  I(X_V;X_W) &:=& H(X_V) + H(X_W) - H(X_{V\cup W})\\
  I(X_V;X_W| X_U) &:= &H(X_{U\cup V}) + H(X_{U\cup W}) \\
   &&{} - H(X_{U\cup V\cup W}) - H(X_{U}).
   \end{array}
 \]  
We extend this notation to almost entropic points.

\section{Copy Lemma}

The following lemma was used (somewhat implicitly) in the very first proof of a non-Shannon type inequality in \cite{zy98},
see also \cite[Lemma~14.8]{yeung-book}. The general  version of  this lemma appeared in \cite{dougherty-six} and \cite{dougherty}.
In \cite{dougherty} it was presented explicitly, with a  complete and detailed proof.

\begin{lemma}[\textbf{Copy Lemma}]
\label{lemma:copy}
For every tuples of jointly distributed random variables $(X,Y,Z)$ there exists a distribution $(A,B,C,C')$ such that
 \begin{itemize}
 \item the distribution $(A,B,C)$ coincides with the distribution of $(X,Y,Z)$; the distribution of $(A,C')$ coincides with the distribution of $(X,Z)$, and
 \item $I(C'; B,C |A) = 0$.
 \end{itemize}
\end{lemma}
\begin{IEEEproof}[Sketch of the proof]
The distribution $(A,B,C,C')$  can be constructed as follows. We start with $A$ that has the same distribution as $X$. Then, for each value of $A$,
we add two  conditional distributions: $(B,C)$ should be distributed (given the value of $A$) like $(Y,Z)$   (given the corresponding
value of $X$), and $C'$ should be distributed (also conditional on  the value of $A$) like $Z$   (given the corresponding value of $X$). 
We may require that  the conditional distributions of $(B,C)$ and $C'$ are independent conditional on $A$. This construction gives the required $(A,B,C,C')$.
\end{IEEEproof}
In what follows we abuse the notation and identify $(A,B,C)$ with $(X,Y,Z)$. That is, we say that the initial distribution $(X,Y,Z)$
can be extended\footnote{Formally speaking, such an ``extension'' may require a subdivision of the initial probabilistic space,
so technically $(A,B,C)$ and  $(X,Y,Z)$ are defined on different probabilistic spaces.} 
to the distribution $(X,Y,Z,Z')$, where  $(X,Z)$ and $(X,Z')$ have isomorphic joint distributions, 
and $I(Z'; Y,Z | X) = 0$. Following \cite{dougherty}, we 
say that $Z'$ is a $Y$-copy of $Z$ over $X$, denoted by
\[
  Z' := Y\text{-}\mathrm{copy}(Z | X).
\]
Note that Lemma~\ref{lemma:copy} remains valid if $X,Y,Z$ are not individual variables but tuples of variables. 
For example, 
$(A_1,A_2,B_1,B_2,C_1,C_2)$ can be extended to a tuple $(A_1,A_2,B_1,B_2,C_1,C_2, C'_1, C_2)$ such that
 \begin{itemize}
 \item the quadruples of random variables $(A_1,A_2, C_1, C_2)$  and $(A_1,A_2,B_1,B_2, C'_1, C'_2)$ have isomorphic  distributions;
 \item $I(C_1',C'_2; B_1,B_2,C_1,C_2 |A_1,A_2) = 0$.
 \end{itemize}
We say that $(C'_1,C'_2)$ is a $(B_1,B_2)$-copy of $(C_1,C_2)$ over $(A_1,A_2)$ and denote it 
 \[
  (C_1',C'_2)  := (B_1,B_2)\text{-}\mathrm{copy}(C_1,C_2 | A_1,A_2).
 \]

F.~Mat\'u\v{s} and L.~Csirmaz proposed several extensions and generalizations of the Copy Lemma 
(\emph{polymatroid convolution}, \emph{book extension}, \emph{maximum entropy extension},
see, e.g.,  \cite{matus-adhesive, book-inequality,csirmaz-matus}). However, to the best of our knowledge,
the ``classical'' version of the Copy Lemma is strong enough for all known proofs of non-Shannon type inequalities

\smallskip
\noindent
All known proofs of (unconditional) non-Shannon type information inequalities can be presented in the following style:
we start with a distribution $(X_1,\ldots, X_n)$; using the Copy Lemma we  supplement to this distribution several new 
variable $(Y_1,\ldots, Y_k)$; then we enumerate the basic Shannon's inequalities for the joint distribution $(X_1,\ldots, X_n,Y_1,\ldots, Y_k)$
and show that a combination of these inequalities (together with the constraints from the Copy Lemma) gives the desired
new inequality for $(X_1,\ldots, X_n)$ (the entropy terms with  new variable $Y_j$ should cancel out). We refer the reader to \cite{dougherty}
for many instances of this argument.

\smallskip

In what follows we use a very similar scheme of  proof. The difference is that we superimpose constraints 
(equalities for the coordinates of the entropy profile of $(X_1,\ldots, X_n)$) that are specific for a given particular problem. That is,
in each application we prove a \emph{conditional} non-Shannon type inequality (which is valid only assuming
some linear constraints for the entropy quantities).  Each of these arguments can be theoretically translated in a more
conventional style:  we can at first prove some new unconditional non-Shannon type inequalities, and then combine them with 
the given constraints and deduce the desired corollary.  In our approach,  throughout the paper we do not need to prove  any  unconditional non-Shannon type inequality for the entropy.

\section{Symmetries}

In this section we discuss simple symmetry considerations that help to reduce the dimension
of a linear programs associated with  problems concerning information inequalities (assuming that the problem has a particular symmetric structure). 
A similar technique (action of permutations on multidimensional distributions and their entropy profiles, information inequalities for
distributions with symmetric entropy profiles) was   studied in \cite{qi-yeung}.

Let  $\mathbf{X}^{(n)}=(X_1,\ldots, X_n)$ be a joint distribution of $n$ random variables. 
We denote by  $\vec H(\mathbf{X})$ its \emph{entropy profile}. 
Let $\pi\ :\ [n]\to [n]$ be a permutation of indices (an $n$-\emph{permutation}). 
This permutation induces a natural transformation of the distribution $\mathbf{X}$,
 \[
  \mathbf{X} = (X_1,\ldots, X_n)\overset{\pi}{\longmapsto} \mathbf{\hat X} := (X_{\pi(1)},\ldots, X_{\pi(n)}),
 \]
 and therefore a transformation of the entropy profile 
  \[
 \vec H(\mathbf{X})  \overset{\pi}{\longmapsto}\vec H(\mathbf{\hat X}).
 \]
 \begin{example}
 If $\pi$ is the transposition of indices $1$ and $2$, then the mapping defined above
exchanges the entropy values $H(X_1)$ and $H(X_2)$, the values  $H(X_1,X_3)$ and $H(X_2,X_3)$, etc., 
and does not change the values $H(X_1,X_2)$, $H(X_1,X_2,X_3)$, etc.
 \end{example}
 
 \smallskip
 
Thus, every permutation of indices $\pi$ induces a transformation of the set of all entropic points $\Gamma_n^*$. 
This transformation of the entropic points can be naturally extended to a linear transformation of the entire space  $V_n$  
(defined by a suitable permutation of the coordinate axis). In what follows we denote this transformation by $T_{\pi}$. 
(Note that $\pi$ is a permutation on $n$ elements, while $T_{\pi}$ is a transformation of  the space  of dimension $2^n-1$.)
In the usual algebraic  terminology,  we can say that we defined a  representation  of the symmetric group 
(the group of all $n$-permutations) by linear transformations of the space $V_n$.

We say that a point $\mathbf{x}\in V_n$  is $\pi$-invariant if  $T_{\pi}(\mathbf{x}) = \mathbf{x}$.  
Similarly, we say that  a set of points $S\subset V_n$ is $\pi$-invariant, if $T_{\pi}(\mathbf{x})$ is a permutation of
points in $S$.  For a subgroup $G$ of $n$-permutations, we say that  a point (or a subset) in $V_n$
is $G$-invariant, if it is  $\pi$-invariant for every $\pi \in G$.  

We denote the set of $G$-invariant points in $V_n$ by $\mathrm{Inv}(G)$.
In what follows we typically study linear and affine spaces that are $G$-invariant for some specific group
of permutations $G$. Note that every $G$-invariant linear space in $V_n$ contains the set
$\mathrm{Inv}(G)$ as a subspace.

The sets of all entropic and all almost entropic points ($\Gamma^*_n$ and $\bar \Gamma_n^*$ respectively) 
are obviously $\pi$-invariant for every $n$-permutation $\pi$. For every  group $G$ of $n$-permutations,
the intersections of $\Gamma^*_n$ or $\bar \Gamma_n^*$ with any  $G$-invariant set gives another $G$-invariant set. 
In what follows we discuss several simple examples of this type.
\begin{example} \label{ex:3}
Let $L_{ent}$ be the set of points in $\Gamma_4^*$ such that 
\[
\left\{
 \begin{array}{rclrcl}
 H(X_1|X_2,X_3)&=&0,&
 H(X_1|X_2,X_4)&=&0,\\
 H(X_2|X_1,X_3)&=&0,&
 H(X_2|X_1,X_4)&=&0,\\
 H(X_1,X_2,X_3,X_4)&=&1.
 \end{array}
 \right.
\]
Since $L_{ent}$ is defined by affine conditions, it can be represented as the intersection of $\Gamma_4^*$
with an affine subspace $L$ (of co-dimension $5$) in the entire space $V_4=\mathbb{R}^{15}$. 
It is easy to see that this subspace $L$ is invariant with respect to
transpositions $\pi_{12}$ and $\pi_{34}$ 
($\pi_{12}$ exchanges the indices $1$ and $2$, and $\pi_{12}$ exchanges the indices  $3$ and $4$).
Therefore, this subspace is $G$-invariant, where $G$ is the subgroup of the symmetric group generated by 
the transpositions $\pi_{12}$ and $\pi_{34}$  (this subgroup consists of four permutations).
\end{example}

\smallskip

We say that a linear functional
 \[
 {\cal L} : \mathbb{R}^{2^n-1} \to \mathbb{R}
 \]
is $\pi$-invariant, if ${\cal L}(T_{\pi}(\mathbf{x})) = {\cal L}(\mathbf{x})$ for all $\mathbf{x}\in  \mathbb{R}^{2^n-1}$. 
Similarly, for a subgroup  of $n$-permutations $G$, we say that a linear functional is $G$-invariant, if it is 
$\pi$-invariant for every $\pi \in G$.

\begin{example} \label{ex:2}
Ingleton's quantity $Ing(X_1,X_2,X_3,X_4)$ defined as 
\[
  I(X_1;X_2|X_3) +  I(X_1;X_2|X_4) + I(X_3:X_4) - I(X_1;X_2)
\]
can be extended to a linear functional on the space $V_4=\mathbb{R}^{15}$. This functional is invariant 
with respect to the transposition of indices $1$ and $2$,  to the transposition of indices $3$ and $4$, and 
with respect to the group (of size four) generated by these two transpositions.
\end{example}

\begin{claim}\label{claim:mass-center}
Let $G=\{\pi_1,\ldots,\pi_k\}$ be a subgroup of $n$-permutations, and let $\mathbf{x}$ be a point in $V_n$. 
Then the center of mass of the points
\begin{equation}
 T_{\pi_1}(\mathbf{x}),\ldots,  T_{\pi_k}(\mathbf{x}) \label{eq:k-points}
 \end{equation}
 is $G$-invariant.
\end{claim}
\begin{IEEEproof}
It is easy to see that the set of points \eqref{eq:k-points} is $\pi$-invariant for each $\pi\in G$. Therefore, 
the center of mass of this set is also $G$-invariant.  
\end{IEEEproof}

\begin{lemma}\label{lemma:inv}
Let $G=\{\pi_1,\ldots,\pi_k\}$  be a subgroup of $n$-permutations, $L$ be a $G$-invariant convex set in $V_n$, 
and ${\cal F} :V_n \to \mathbb{R}$ be a $G$-invariant linear function.
Then
 \[
  \inf\limits_{\mathbf{x}\in L}  {\cal F} (\mathbf{x}) =   \inf\limits_{\mathbf{y}\in L\cap \mathrm{Inv}(G)}  {\cal F} (\mathbf{y}).
 \]
\end{lemma}
\begin{IEEEproof}
It is enough to prove that for every $\mathbf{x}\in L$ there exists a $\mathbf{y}\in L\cap \mathrm{Inv}(G)$ such that
 \[
{\cal F} (\mathbf{x}) = {\cal F} (\mathbf{y}).
 \]
To this end we take the points  
 \[
   T_{\pi_1}(\mathbf{x}), \ldots,  T_{\pi_k}(\mathbf{x})
 \]
and define $y$ as the center of mass of these $k$ points. Since  $L$ is $G$-invariant, each point $T_{\pi_i}(\mathbf{x})$ belongs to $L$.
From convexity of $L$ it follows that  $\mathbf{y}$ also belongs to $L$, and Claim~\ref{claim:mass-center} implies that $\mathbf{y}$
belongs to $ \mathrm{Inv}(G)$. Thus, the constructed point belongs to $ L\cap \mathrm{Inv}(G)$.

On the other hand, from $G$-invariance of $\cal F$ it follows that 
\begin{equation}\label{eq:k-points-equal}
   {\cal F}(T_{\pi_1}(\mathbf{x})) =  \ldots =    {\cal F}(T_{\pi_k}(\mathbf{x})).
\end{equation}
Since ${\cal F}$ is linear, we conclude that ${\cal F}$ at the center of mass of $ T_{\pi_i}(\mathbf{x})$
 is equal to the same values as the points in  \eqref{eq:k-points-equal}, and the lemma is proven.
\end{IEEEproof}
\begin{example}
Assume we are looking for the minimum of $Ing(X_1,X_2,X_3,X_4)$ (see Example~\ref{ex:2}) on the set of \emph{almost entropic} 
points (points in $\bar \Gamma_4^*$) that satisfy the linear constraints
\[
\left\{
 \begin{array}{rclrcl}
 H(X_1|X_2,X_3)&=&0,&
 H(X_1|X_2,X_4)&=&0,\\
 H(X_2|X_1,X_3)&=&0,&
 H(X_2|X_1,X_4)&=&0,\\
 H(X_1,X_2,X_3,X_4)&=&1.
 \end{array}
 \right.
\]
 (see Example~\ref{ex:3}). We define $G$ as the group of permutations generated by two
 transpositions  $\pi_{12}$ and $\pi_{34}$ that exchange the indices $(1,2)$ and $(3,4)$ respectively.
 
We claim that the required extremal value of  $Ing(X_1,X_2,X_3,X_4)$
 is achieved at some $G$-invariant point.  
 Indeed, the given constraints define in $V_4$ a $G$-invariant affine subspace. The intersection of this subspace 
 with the convex cone $\bar \Gamma_4^*$ is a  $G$-invariant convex set.
On the other hand,
 Ingleton's quantity  $Ing(X_1,X_2,X_3,X_4)$ is a  $G$-invariant linear function. 
 Therefore, we can apply Lemma~\ref{lemma:inv} and conclude that desired extremum is achieved 
 at some point in $\mathrm{Inv}(G)$.
 \end{example}

\section{The Standard Benchmark: Bounds for Ingleton's quantity}

We start with a warm-up: in this section we show how to prove in terms of direct applications of 
the Copy Lemma the bound for Ingleton's inequality from \cite{dougherty}, and how symmetry considerations
can be used in arguments of this type.

It is widely believed that one of the most interesting questions on the structure of $\bar \Gamma_4^*$
is the question on the Ingleton's quantity (see Example~\ref{ex:2} above): what is the worst possible violation of Ingleton's' inequality
\begin{equation}\label{eq:ingleton}
Ing(A,B,C,D) \ge 0.
\end{equation}
This inequality is true for \emph{linearly representable} entropic points  but 
for some other entropic points it is violated (see \cite{hrsv}). The question is how far below $0$ can go the \emph{Ingleton score}
\[
\frac{Ing(A,B,C,D)}{H(A,B,C,D)}
\]
(here we use the terminology from \cite{dougherty}). From the Shannon type inequalities it follows only that this score is greater than
$-1/4$. The first non-Shannon type inequality from \cite{zy98} implies that this score is greater than
$-1/6$. The best known bound was discovered  by Dougherty  \emph{et al.}:
\begin{theorem}[originally proven in \cite{dougherty}]\label{th:ingleton}
For every quadruple of  jointly distributed random variables $(A,B,C,D)$
\[
\frac{Ing(A,B,C,D)}{H(A,B,C,D)} \ge - {3}/{19}.
\]
\end{theorem}
(To the best of our knowledge, the  best published upper bound on the infimal Ingleton score is $-0.09243$,  \cite{csirmaz-matus},
 which refuted the  four-atom conjecture from \cite{dougherty}.)

We show that Theorem~\ref{th:ingleton} can be proven with only three applications of the Copy Lemma
(with four new variables), with the help of symmetry considerations from the previous section.
This approach gives a pretty simple way to  confirm the ratio $-3/19\approx -0.15789$
with the help of a computer, and provides a reasonably short ``formally checkable'' proof
(a linear combination of 167 simple equalities and inequalities with rational coefficients, see Appendix).
\begin{IEEEproof}[Sketch of the proof]
Our goal it to find the minimal value of the Ingleton score \eqref{eq:ingleton} for entropic points. 
It is equivalent to the minimal value of $Ing(A,B,X,Y)$ for \emph{almost} entropic points satisfying
the normalization condition $H(A,B,C,D)=1$.  

The objective function $Ing(A,B,C,D)$ and the normalization constraint are invariant with respect to 
the transpositions $\pi_{A,B}$ and $\pi_{C,D}$ that exchange the variables $A,B$ and $C,D$ respectively.
Therefore, we can apply Lemma~\ref{lemma:inv} and conclude that the required minimal value can be found
in the space of points that are  $\pi_{A,B}$- and $\pi_{C,D}$-invariant.

The invariance means that we can restrict ourselves to the points in $\bar \Gamma_4^*$ such that  
\begin{equation}\label{eq:ingleton-sym}
\left\{
\begin{array}{l}
H(A) = H(B),\ H(A,C,D) = H(B,C,D),\\
H(A,C) = H(B,C),\
H(A,D) = H(B,D),\\
H(C) = H(D),\ H(A,B,C) = H(A,B,D),\\
H(A,C) = H(A,D),\
H(B,C) = H(B,D).\\

\end{array}
\right.
\end{equation}
The crucial part of the proof is, of course, the Copy Lemma. We use the trick\footnote{%
It is remarkable that Ineq.~33. from \cite{dougherty} 
(even together with the symmetry constraints)
does not imply Theorem~\ref{th:ingleton}. There is no contradiction: the  instances of the Copy Lemma 
used in this proof  imply  besides Ineq.~33. several other non-Shannon type inequalities,
and only together these inequalities  imply the required bound for the Ingleton score.
} 
from \cite[proof of Ineq.~33, p.~19]{dougherty}
and define four new random variables $R,S,T,U$:
\begin{equation}\label{eq:ingleton-copy-lemma}
\left\{
\begin{array}{ccl}
(R,S) &: =& \emptyset\text{-}\mathrm{copy}(B,D | A,C), \\
T &: =& (D,R)\text{-}\mathrm{copy}(C|A,B,S),\\
U &: =& D\text{-}\mathrm{copy}(B|A,C,R,S,T).
\end{array}
\right.
\end{equation}
Technically, this means that we take the conditions (equalities)  from the Copy Lemma (Lemma~\ref{lemma:copy}) 
that determine the entropies of the introduced variables and the properties of conditional independence for each instance of the Copy Lemma.

We combine  the constraints  \eqref{eq:ingleton-sym} and \eqref{eq:ingleton-copy-lemma} with the normalization condition $H(A,B,C,D)=1$ and
all Shannon type inequalities for $A,B,C,D,R,S,T,U$.  Thus, we obtain a linear program for the $2^8-1=255$ variables (corresponding to the coordinates of the the entropy profile of  $(A,B,C,D,R,S,T,U)$) with the objective function
 \[
 Ing(A,B,C,D) \to \min
 \]
This problem is easy  for the standard linear programming solvers 
(we made experiments with the solvers \cite{qsopt,glpk,gurobi}). 
We can extract from the dual solution of this linear program a rational combination of inequalities that gives the desired
bound (without  floating point arithmetics nor rounding). We present this  ``formally checkable'' proof in Appendix.
\end{IEEEproof}

\section{Secret Sharing Schemes on the V\'amos Matroid}

The notion of a \emph{secret sharing scheme} was  introduced by Shamir \cite{shamir} and Blakley \cite{blakley}.
Nowadays,  secret sharing is an important component of many cryptographic protocols. 
We refer the reader to \cite{secret-sharing-survey} for an excellent survey of secret sharing and its applications. 

The aim of  \emph{perfect secret sharing} is to distribute a \emph{secret value}  among a set of parties in such a way that
only the \emph{qualified sets} of parties can recover the secret value; the non-qualified sets should have no information about the secret.
The family of qualified sets of parties is called the \emph{access structure} of the scheme. This family of sets must be monotone
(a superset of a qualified set is also qualified).

The \emph{information ratio} of a secret sharing schemes is the maximal ratio between the size (the entropy) of a share and 
the size (the entropy) of the secret. The optimization of this parameter is the central problem of secret sharing. 
This question was extensively studied for several families of access structures. 
The main technique for proving lower bounds on complexity of secret sharing schemes is nowadays the linear programming
applied to the constraints formulated in terms of  classical and non-classical information inequalities, see \cite{padro-et-al-2010}.

In what follows we apply our technique (the Copy Lemma combined with symmetry considerations and linear programming)
to one of the classical access structures  --- the access structure defined on the V\'amos matroid 
(more technically,  {one} of two access structures that can be defined on the V\'amos matroid).
We skip the general motivation 
(the V\'amos matroid is particularly interesting as a simple example of a non-representable matroid, see \cite{secret-sharing-survey} for details) 
and  formulate directly the problem of the information ratio for this access structure  in terms of Shannon's entropy of the involved random variables.

We define an access structure with parties $\{1,\ldots, 7\}$.
The minimal qualified sets are the $3$-sets $\{1,2,3\}$, $\{1,4,5\}$ and all $4$-sets not containing them, with three exceptions 
$\{2,3,4,5\}$, $\{2,3,6,7\}$, $\{4,5,6,7\}$. We deal with jointly distributed random variables  $(S_0,\ldots, S_7)$,
where $S_0$ is the secret and $S_1,\ldots,S_7$ are the shares given to the seven parties involved in the scheme.
We require that 
\begin{itemize}
\item[(i)] for every minimal qualified set $V$ listed above $H(S_0 | S_V) = 0$;
\item[(ii)] if $V\subset\{1,\ldots,7\}$ includes no minimal qualified set, then $H(S_0 | S_V) = H(S_0)$.
\end{itemize}
Under these constraints, we are looking for the extremal value
\[
\frac{\max \{ H(S_1),\ldots, H(S_7) \} }{H(S_0)} \to \inf
\]
Since the set of almost-entropic points is a closed convex cone, we can add the normalization $H(S_0)=1$ and rewrite
the objective function as
\begin{equation}\label{eq:V\'amos-obj}
\max \{ H(S_1),\ldots, H(S_7) \} \to \min
\end{equation}
It is known that the required minimum is not greater than $4/3$, \cite{vamos-upper-bound}. The history of improvements of the 
lower bound for this problem is presented in the table:

\medskip

\begin{tabular}{r | l}
\cite{seymour}, 1992 & $>1$ \\
\cite{beimel-livne}, 2006 & $\ge 1+\Omega(1/\sqrt{k})$ for a secret of size $k$\\
\cite{beimel-livne-padro}, 2008 & $\ge11/10$ \\
\cite{metcalf-burton}, 2011 & $\ge9/8=1.125$ \\
\cite{hadian-2013}, 2013 &  $\ge 67/59\approx 1.135593$ \\ 
\cite{tarik2018}, 2018 & $\ge 33/29 \approx 1.137931$ \\ 
this paper & $\ge 561/491\approx 1.142566$ \\
\end{tabular}

\medskip

Notice that the objective function, the normalization condition $H(S_0)=1$, 
and the linear constraints in (i) and (ii) above are invariant with respect to the permutation that
 \begin{itemize}
 \item swap the indices $2$ and $3$,
 \item swap the indices $4$ and $5$,
 \item swap the indices $6$ and $7$,
 \item swap the pairs $(2,3)$ and $(4,5)$.
 \end{itemize}
Therefore we can apply Lemma~\ref{lemma:inv} and reduce the problem to those almost  entropic points that 
satisfy the symmetries defined above.

We denote 
\[
V:=(S_0,S_1)\text{ and }W:=(S_6,S_7)
\]
and define four new variables by applying twice the Copy Lemma:
\begin{equation}
\nonumber
\begin{array}{l}
\text{let }(V',W') \text{ be an } (S_0,S_1,S_6,S_7)\text{-copy of }(V,W) \\
 \rule{18mm}{0mm} \text{ over }(S_2,S_3,S_4,S_5); \\ 
\text{let }(V'',W'') \text{ be an } (S_0,S_1,S_6,S_7,V',W')\text{-copy of }(V,W) \\
 \rule{18mm}{0mm}  \text{ over }(S_2,S_3,S_4,S_5).
\end{array}
\end{equation}

\begin{remark}
Each instance of the Copy Lemma includes the property of \emph{conditional independence}.
In the construction described above we have two applications of the Copy Lemma, and
therefore two  independence conditions:
 \[
 \begin{array}{l}
  I(V',W' ; S_0,S_1,S_6,S_7 | S_2,S_3,S_4,S_5) = 0\\
  I(V'',W'' ; S_0,S_1,S_6,S_7,V',W' | S_2,S_3,S_4,S_5) = 0.
\end{array}
 \]
These two conditions can be merged in one (more symmetric)
constraint
\begin{equation}\nonumber
\begin{array}{l}
H(S_0,S_1,S_6,S_7,V',V'',W',W''|S_2,S_3,S_4,S_5) = {} \\
\rule{5mm}{0mm} {} = H(S_0,S_1,S_6,S_7|S_2,S_3,S_4,S_5) \\
\rule{20mm}{0mm}  {} + H(V',W'|S_2,S_3,S_4,S_5) \\
\rule{25mm}{0mm} {} + H(V'',W''|S_2,S_3,S_4,S_5).
\end{array}
\end{equation}
\end{remark}
Thus, we obtain a linear program with the following constraints:
 \begin{itemize}
 \item the conditions (i) and (ii) that define the access structure of the secret sharing scheme,
 \item the normalization $H(S_0)=1$,
 \item the equalities that follow from the symmetry of the access structure,
 \item the conditions of the two instances of the Copy Lemma,
 \item and the Shannon type inequalities for twelve random variables $(S_0,\ldots,S_7, V',V'',W',W'')$
 (i.e., for $2^{12}-1$ coordinates of their entropy profile).
 \end{itemize}
 The goal is to optimize the objective function \eqref{eq:V\'amos-obj}.
 
  \smallskip
 
 \begin{remark}
 We do not need to include in the linear program the variables $V$ and $W$ and their entropies ---
 these two variables are used only as a notation in the Copy Lemma. 
 That is, our a linear program includes  $2^{12}-1$  real-valued variables (but  not  $2^{14}-1$).
 \end{remark}
 
 \smallskip
 
With the help of a computer (we used the solvers \cite{qsopt,glpk,gurobi}) we find the  optimal solution of this linear program: 
it is equal to $561/491$. Thus we establish the following statement.
\begin{theorem}\label{th:vamos}
The optimal information rate of the secret sharing scheme for the access structure $V_0$ on the V\'amos matroid is not less
than $561/491 =  1.142566\ldots$
\end{theorem}
\begin{remark}
The standard linear programming solvers use the floating point arithmetic, and we should keep in mind  the rounding errors. 
Our linear program contains only constraints with integer coefficients, 
so  the values of the variables  for the optimal solution of this linear program (the primal and the dual) can be taken rational. 
These rational values can be found by exact computing, without rounding. 
To compute the exact solution of a linear program we use the rational linear programming solver \texttt{QSopt\_ex}, see  \cite{esponoza-phd,qsopt}. 

The found exact dual solutions (rational linear combinations of constraints)  for this linear program
consist of more than $10^3$ equalities and inequalities. The found rational linear combinations of  equalities and inequalities 
provide a formal mathematical proof of Theorem~\ref{th:vamos},  though such a  proof hardly can be called ``human-readable.''

Computing the exact solution of  a linear program is rather time consuming. 
If we only need a \emph{lower bound} for the optimal solution, it is enough to find a feasible solution of the dual problem that  
approximates (well enough) the optimal solution. Such an approximation can be found much faster than the exact rational solution,
with more conventional floating point LP solvers.
\end{remark}
  \begin{remark} 
 The linear program constructed in this section would have the same optimal value if we omit the symmetry conditions
 (note that, unlike the previous section,  in the proof of Theorem~\ref{th:vamos} the Copy Lemma is applied in symmetric settings). However,  these symmetry conditions are not useless.
 In our experiments with exact computations, the symmetry constraints help to  find a rational solution of the dual problem with smaller 
 denominators. Also, in several experiments with floating point solvers, the symmetry constraints make  the computation (or approximation) of
 the optimal solution faster. However, this is not a general rule:  in some experiments  the symmetry constraints make the computation even slower --- it depends on  the used  linear programming algorithm and on the chosen scaling of the objective function and of the constraints. 
  \end{remark}

\section{Conclusion and perspectives}

 \paragraph{From the Copy Lemma towards structural properties of distributions}
When we derive balanced information inequalities,  the technique of the Copy Lemma with ``clones'' of \emph{individual} random variables
 is equivalent to the technique of  the Ahlswede--K\"orner lemma, see \cite{tarik-equiv}.
In this paper we used the Copy Lemma in a stronger form, making ``clones'' of \emph{pairs} of correlated variables.
In the proofs of  Theorem~\ref{th:ingleton} and Theorem~\ref{th:vamos} it was crucial that we can ``duplicate'' in one shot
a pair of correlated random variables,  not only a single random variable.

The Ahlswede--K\"orner lemma used in \cite{tarik2018} has a clear intuitive meaning:
it can be interpreted in  terms  of extraction of the \emph{common information}, see \cite{ak-lemma,ak-lemma-ineq}. 
It would be interesting to reveal similar structural properties of the probability distribution that 
would give an  intuitive explanation of the efficiency of the Copy Lemma  with \emph{two} (or more) copied variables. 

\smallskip

 \paragraph{General automorphisms}
In this paper we  used  only  very simple  symmetries of distributions generated by swapping individual random variables or pairs of random variables. 
However, the symmetry considerations  (see Lemma~\ref{lemma:inv}) apply in a much more general setting.
Let us mention that  several non-trivial symmetries was used in  \cite{apte-walsh}  to reduce the size of a linear program
 corresponding to a network coding problem.
An anonymous referee suggested a natural direction of research: to study  problems in Shannon's theory (on  network coding, secret sharing, etc.) 
where  the size of a relevant linear program can be reduced with the help of more complex groups of automorphisms.

\smallskip

 \paragraph*{Acknowledgment}
 This  work  was  supported  in part by  ANR  under  grant   ANR-15-CE40-0016-01 RaCAF.

\bigskip

\IEEEtriggeratref{25}



\newpage

\appendix

\small

\smallskip

\noindent
In what follows we present a list of $167$ equalities and inequalities. An integer linear  combination of these inequalities  implies 
Theorem~\ref{th:ingleton}.

\bigskip

\tiny

\noindent
\textbf{Definition of the standard information quantities:}
\smallskip

\noindent with a factor of $4503$: \\  $ 
H(A) + H(B) - H(A,B) - I(A;B) = 0$

\noindent with a factor of $-4503$: \\ $ 
H(AC) + H(B,C) - H(A,B,C) - H(C) - I(A;B|C) = 0$

\noindent with a factor of $-4503$: \\ $ 
H(A,D) + H(B,D) - H(A,B,D) - H(D) - I(A;B|D) = 0$

\noindent with a factor of $-4503$: \\ $ 
H(C) + H(D) - H(C,D) - I(C;D) = 0$


\smallskip
\noindent
\textbf{Symmetries between $A$ and $B$ and between $C$ and $D$:}
\smallskip

\noindent with a factor of $948$: \\ $ 
 H(A) - H(B) = 0$

\noindent with a factor of $-2370$: \\ $ 
 H(A,C) - H(B,C) = 0$

\noindent with a factor of $1389$: \\ $ 
 H(A,C,D) - H(B,C,D) = 0$

\noindent with a factor of $711$: \\ $ 
 H(C) - H(D) = 0$

\noindent with a factor of $-2844$: \\ $ 
 H(B,C) - H(B,D) = 0$

\noindent with a factor of $1859$: \\ $ 
 H(A,B,C) - H(A,B,D) = 0$
 
 \smallskip
\noindent
\textbf{Conditions from the Copy Lemma:}
\smallskip

\noindent with a factor of $-5260$: \\ $ 
H(A,C,R,S) + H(A,B,C,D) - H(A,B,C,D,R,S) - H(A,C) = 0$

\noindent with a factor of $948$: \\ $ 
H(R) - H(B) = 0$

\noindent with a factor of $948$: \\ $ 
H(R,S) - H(B,D) = 0$

\noindent with a factor of $-948$: \\ $ 
H(A,R) - H(A,B) = 0$

\noindent with a factor of $-3792$: \\ $ 
H(A,S) - H(A,D) = 0$

\noindent with a factor of $948$: \\ $ 
H(A,R,S) - H(A,B,D) = 0$

\noindent with a factor of $-948$: \\ $ 
H(C,R) - H(B,C) = 0$

\noindent with a factor of $2873$: \\ $ 
H(C,S) - H(C,D) = 0$

\noindent with a factor of $-1925$: \\ $ 
H(C,R,S) - H(B,C,D) = 0$

\noindent with a factor of $1385$: \\ $ 
H(A,C,R) - H(A,B,C) = 0$

\noindent with a factor of $919$: \\ $ 
H(A,C,S) - H(A,C,D) = 0$

\noindent with a factor of $1488$: \\ $
H(A,C,R,S) - H(A,B,C,D) = 0$
 
\noindent with a factor of $-3555$: \\ $ 
 H(A,B,S,T) + H(A,B,C,D,R,S) - H(A,B,C,D,R,S,T) - H(A,B,S) = 0$

\noindent with a factor of $474$: \\ $ 
 H(T) - H(C) = 0$

\noindent with a factor of $948$: \\ $ 
 H(S,T) - H(C,S) = 0$

\noindent with a factor of $-1613$: \\ $ 
 H(A,T) - H(A,C) = 0$

\noindent with a factor of $-2133$: \\ $ 
 H(B,T) - H(B,C) = 0$

\noindent with a factor of $-520$: \\ $ 
 H(A,S,T) - H(A,C,S) = 0$

\noindent with a factor of $-46$: \\ $ 
 H(A,B,T) - H(A,B,C) = 0$

\noindent with a factor of $2890$: \\ $ 
 H(A,B,S,T) - H(A,B,C,S) = 0$

\noindent with a factor of $-3855$: \\ $ 
 H(A,C,R,S,T,U) + H(A,B,C,D,R,S,T) - H(A,B,C,D,R,S,T,U) - H(A,C,R,S,T) = 0$

\noindent with a factor of $1896$: \\ $ 
 H(U) - H(B) = 0$
 
\noindent with a factor of $-1659$: \\ $ 
 H(T,U) - H(B,T) = 0$

\noindent with a factor of $46$: \\ $ 
 H(A,U) - H(A,B) = 0$

\noindent with a factor of $-1555$: \\ $ 
 H(C,U) - H(B,C) = 0$

\noindent with a factor of $341$: \\ $ 
 H(R,T,U) - H(B,R,T) = 0$

\noindent with a factor of $-46$: \\ $ 
 H(A,T,U) - H(A,B,T) = 0$

\noindent with a factor of $607$: \\ $ 
 H(C,T,U) - H(B,C,T) = 0$

\noindent with a factor of $-341$: \\ $ 
 H(A,R,U) - H(A,B,R) = 0$

\noindent with a factor of $-341$: \\ $ 
 H(C,R,U) - H(B,C,R) = 0$

\noindent with a factor of $-46$: \\ $ 
 H(A,S,U) - H(A,B,S) = 0$

\noindent with a factor of $46$: \\ $ 
 H(C,S,U) - H(B,C,S) = 0$

\noindent with a factor of $191$: \\ $
 H(A,C,U) - H(A,B,C) = 0$

\noindent with a factor of $-46$: \\ $ 
 H(C,R,S,U) - H(B,C,R,S) = 0$

\noindent with a factor of $-667$: \\ $ 
 H(A,C,S,U) - H(A,B,C,S) = 0$

\noindent with a factor of $-428$: \\ $ 
 H(A,S,T,U) - H(A,B,S,T) = 0$

\noindent with a factor of $-266$: \\ $ 
 H(A,C,T,U) - H(A,B,C,T) = 0$

\noindent with a factor of $817$: \\ $ 
 H(A,C,R,U) - H(A,B,C,R) = 0$

\noindent with a factor of $114$: \\ $ 
 H(A,C,S,T,U) - H(A,B,C,S,T) = 0$

\noindent with a factor of $1337$: \\ $ 
 H(A,C,R,S,T,U) - H(A,B,C,R,S,T) = 0$


\smallskip
\noindent
\textbf{Shannon type inequalities:}
\smallskip

\noindent with a factor of $711$: \\ $ 
H(C) + H(D) - H(C,D) \ge 0$

\noindent with a factor of $549$: \\ $ 
H(A,B,C,D,S,U) - H(B,C,D,S,U) \ge 0$

\noindent with a factor of $195$: \\ $ 
H(A,B,C,D,S,T,U) - H(B,C,D,S,T,U) \ge 0$

\noindent with a factor of $678$: \\ $ 
H(A,B,C,D,S,T,U) - H(A,C,D,S,T,U) \ge 0$

\noindent with a factor of $136$: \\ $ 
H(A,B,C,D,S) - H(A,B,D,S) \ge 0$

\noindent with a factor of $849$: \\ $ 
H(A,B,C,D,S,T,U) - H(A,B,D,S,T,U) \ge 0$

\noindent with a factor of $437$: \\ $ 
H(A,C,D,R,S,U) - H(A,C,D,R,U) \ge 0$

\noindent with a factor of $341$: \\ $ 
H(A,R,T) + H(B,R,T) - H(A,B,R,T) - H(R,T) \ge 0$

\noindent with a factor of $46$: \\ $ 
H(A,D,U) + H(B,D,U) - H(A,B,D,U) - H(D,U) \ge 0$

\noindent with a factor of $474$: \\ $ 
H(A,D,T) + H(B,D,T) - H(A,B,D,T) - H(D,T) \ge 0$

\noindent with a factor of $191$: \\ $ 
H(A,D,S,U) + H(B,D,S,U) - H(A,B,D,S,U) - H(D,S,U) \ge 0$

\noindent with a factor of $370$: \\ $
H(A,C,T) + H(B,C,T) - H(A,B,C,T) - H(C,T) \ge 0$

\noindent with a factor of $208$: \\ $ 
H(A,C,D,R,S,U) + H(B,C,D,R,S,U) - H(A,B,C,D,R,S,U) - H(C,D,R,S,U) \ge 0$

\noindent with a factor of $341$: \\ $ 
H(A,R,U) + H(C,R,U) - H(A,C,R,U) - H(R,U) \ge 0$

\noindent with a factor of $607$: \\ $ 
H(A,R,T) + H(C,R,T) - H(A,C,R,T) - H(R,T) \ge 0$

\noindent with a factor of $191$: \\ $ 
H(A,B,S,T,U) + H(B,C,S,T,U) - H(A,B,C,S,T,U) - H(B,S,T,U) \ge 0$

\noindent with a factor of $46$: \\ $ 
H(A,B,R,S,U) + H(B,C,R,S,U) - H(A,B,C,R,S,U) - H(B,R,S,U) \ge 0$

\noindent with a factor of $948$: \\ $ 
H(A,S,T) + H(R,S,T) - H(A,R,S,T) - H(S,T) \ge 0$

\noindent with a factor of $208$: \\ $ 
H(A,C,D,S) + H(C,D,R,S) - H(A,C,D,R,S) - H(C,D,S) \ge 0$

\noindent with a factor of $46$: \\ $ 
H(A,B,S,U) + H(B,R,S,U) - H(A,B,R,S,U) - H(B,S,U) \ge 0$

\noindent with a factor of $744$: \\ $ 
H(A,B,C,D) + H(B,C,D,R) - H(A,B,C,D,R) - H(B,C,D) \ge 0$

\noindent with a factor of $927$: \\ $
H(A,B,C,D,S) + H(B,C,D,R,S) - H(A,B,C,D,R,S) - H(B,C,D,S) \ge 0$

\noindent with a factor of $744$: \\ $ 
H(A,B,C,D,R,T) + H(B,C,D,R,S,T) - H(A,B,C,D,R,S,T) - H(B,C,D,R,T) \ge 0$

\noindent with a factor of $266$: \\ $ 
H(A,D,U) + H(D,T,U) - H(A,D,T,U) - H(D,U) \ge 0$

\noindent with a factor of $237$: \\ $ 
H(A,B,C,D,U) + H(B,C,D,T,U) - H(A,B,C,D,T,U) - H(B,C,D,U) \ge 0$

\noindent with a factor of $403$: \\ $ 
H(A,B,C,D,R) + H(B,C,D,R,T) - H(A,B,C,D,R,T) - H(B,C,D,R) \ge 0$

\noindent with a factor of $100$: \\ $ 
H(A,C,D) + H(C,D,U) - H(A,C,D,U) - H(C,D) \ge 0$

\noindent with a factor of $208$: \\ $ 
H(A,C,D,S,T) + H(C,D,S,T,U) - H(A,C,D,S,T,U) - H(C,D,S,T) \ge 0$

\noindent with a factor of $208$: \\ $ 
H(A,C,D,R,S) + H(C,D,R,S,U) - H(A,C,D,R,S,U) - H(C,D,R,S) \ge 0$

\noindent with a factor of $744$: \\ $ 
H(A,B,C,D,S) + H(B,C,D,S,U) - H(A,B,C,D,S,U) - H(B,C,D,S) \ge 0$

\noindent with a factor of $195$: \\ $ 
H(A,B,C,D,R,S,T) + H(B,C,D,R,S,T,U) - H(A,B,C,D,R,S,T,U) - H(B,C,D,R,S,T) \ge 0$

\noindent with a factor of $341$: \\ $ 
H(B,C) + H(C,R) - H(B,C,R) - H(C) \ge 0$

\noindent with a factor of $1925$: \\ $ 
H(B,C,S) + H(C,R,S) - H(B,C,R,S) - H(C,S) \ge 0$

\noindent with a factor of $46$: \\ $ 
H(B,C,S,U) + H(C,R,S,U) - H(B,C,R,S,U) - H(C,S,U) \ge 0$

\noindent with a factor of $403$: \\ $ 
H(A,B,C,D,U) + H(A,C,D,R,U) - H(A,B,C,D,R,U) - H(A,C,D,U) \ge 0$

\noindent with a factor of $778$: \\ $ 
H(A,B,C,D,S) + H(A,C,D,R,S) - H(A,B,C,D,R,S) - H(A,C,D,S) \ge 0$

\noindent with a factor of $191$: \\ $ 
H(B,D,U) + H(D,S,U) - H(B,D,S,U) - H(D,U) \ge 0$

\noindent with a factor of $208$: \\ $ 
H(B,C,D,T) + H(C,D,S,T) - H(B,C,D,S,T) - H(C,D,T) \ge 0$

\noindent with a factor of $3601$: \\ $ 
H(A,B) + H(A,S) - H(A,B,S) - H(A) \ge 0$

\noindent with a factor of $46$: \\ $ 
H(A,B,U) + H(A,S,U) - H(A,B,S,U) - H(A,U) \ge 0$

\noindent with a factor of $476$: \\ $ 
H(A,B,C,R,U) + H(A,C,R,S,U) - H(A,B,C,R,S,U) - H(A,C,R,U) \ge 0$

\noindent with a factor of $362$: \\ $ 
H(A,B,C,D) + H(A,C,D,S) - H(A,B,C,D,S) - H(A,C,D) \ge 0$

\noindent with a factor of $744$: \\ $ 
H(A,B,C,D,R,U) + H(A,C,D,R,S,U) - H(A,B,C,D,R,S,U) - H(A,C,D,R,U) \ge 0$

\noindent with a factor of $208$: \\ $ 
H(B,C,D) + H(C,D,T) - H(B,C,D,T) - H(C,D) \ge 0$

\noindent with a factor of $1181$: \\ $ 
H(A,B,C,D,R,S,U) + H(A,C,D,R,S,T,U) - H(A,B,C,D,R,S,T,U) - H(A,C,D,R,S,U) \ge 0$

\noindent with a factor of $237$: \\ $ 
H(B,C,T) + H(C,T,U) - H(B,C,T,U) - H(C,T) \ge 0$

\noindent with a factor of $403$: \\ $ 
H(B,C,D,R) + H(C,D,R,U) - H(B,C,D,R,U) - H(C,D,R) \ge 0$

\noindent with a factor of $237$: \\ $ 
H(A,B,S,T) + H(A,S,T,U) - H(A,B,S,T,U) - H(A,S,T) \ge 0$

\noindent with a factor of $963$: \\ $ 
H(A,B,C,S,T) + H(A,C,S,T,U) - H(A,B,C,S,T,U) - H(A,C,S,T) \ge 0$

\noindent with a factor of $1337$: \\ $ 
H(A,B,C,R,S,T) + H(A,C,R,S,T,U) - H(A,B,C,R,S,T,U) - H(A,C,R,S,T) \ge 0$

\noindent with a factor of $578$: \\ $ 
H(A,B,C,D,T) + H(A,C,D,T,U) - H(A,B,C,D,T,U) - H(A,C,D,T) \ge 0$

\noindent with a factor of $778$: \\ $ 
H(A,B,C,D,R) + H(A,C,D,R,U) - H(A,B,C,D,R,U) - H(A,C,D,R) \ge 0$

\noindent with a factor of $711$: \\ $ 
H(C,U) + H(D,U) - H(C,D,U) - H(U) \ge 0$

\noindent with a factor of $237$: \\ $ 
H(B,C,U) + H(B,S,U) - H(B,C,S,U) - H(B,U) \ge 0$

\noindent with a factor of $46$: \\ $ 
H(A,B,C,T,U) + H(A,B,S,T,U) - H(A,B,C,S,T,U) - H(A,B,T,U) \ge 0$

\noindent with a factor of $31065_ 
H(A,B,C,D,U) + H(A,B,D,S,U) - H(A,B,C,D,S,U) - H(A,B,D,U) \ge 0$

\noindent with a factor of $849$: \\ $ 
H(A,B,C,D,T) + H(A,B,D,S,T) - H(A,B,C,D,S,T) - H(A,B,D,T) \ge 0$

\noindent with a factor of $80180_ 
H(C,U) + H(T,U) - H(C,T,U) - H(U) \ge 0$

\noindent with a factor of $208$: \\ $ 
H(C,D,U) + H(D,T,U) - H(C,D,T,U) - H(D,U) \ge 0$

\noindent with a factor of $191$: \\ $ 
H(B,C,S,U) + H(B,S,T,U) - H(B,C,S,T,U) - H(B,S,U) \ge 0$

\noindent with a factor of $948$: \\ $ 
H(A,C,R,S) + H(A,R,S,T) - H(A,C,R,S,T) - H(A,R,S) \ge 0$

\noindent with a factor of $312$: \\ $ 
H(A,C,D,U) + H(A,D,T,U) - H(A,C,D,T,U) - H(A,D,U) \ge 0$

\noindent with a factor of $191$: \\ $ 
H(A,C,D,S,U) + H(A,D,S,T,U) - H(A,C,D,S,T,U) - H(A,D,S,U) \ge 0$

\noindent with a factor of $1323$:\\ $ 
H(A,B,C,D) + H(A,B,D,T) - H(A,B,C,D,T) - H(A,B,D) \ge 0$

\noindent with a factor of $46$: \\ $ 
H(A,B,C,D,U) + H(A,B,D,T,U) - H(A,B,C,D,T,U) - H(A,B,D,U) \ge 0$

\noindent with a factor of $341$: \\ $ 
H(A,B,C,D,R) + H(A,B,D,R,T) - H(A,B,C,D,R,T) - H(A,B,D,R) \ge 0$

\noindent with a factor of $198$:\\ $ 
H(A,B,C,D) + H(A,B,D,U) - H(A,B,C,D,U) - H(A,B,D) \ge 0$

\noindent with a factor of $39$: \\ $ 
H(A,B,C,D,R) + H(A,B,D,R,U) - H(A,B,C,D,R,U) - H(A,B,D,R) \ge 0$

\noindent with a factor of $849$: \\ $ 
H(A,B,C,D,R,S,T) + H(A,B,D,R,S,T,U) - H(A,B,C,D,R,S,T,U) - H(A,B,D,R,S,T) \ge 0$

\noindent with a factor of $1879$: \\ $ 
H(B,C,D,S) + H(B,C,R,S) - H(B,C,D,R,S) - H(B,C,S) \ge 0$

\noindent with a factor of $1181$: \\ $ 
H(A,C,D,S,T,U) + H(A,C,R,S,T,U) - H(A,C,D,R,S,T,U) - H(A,C,S,T,U) \ge 0$

\noindent with a factor of $817$: \\ $ 
H(A,B,C,D) + H(A,B,C,R) - H(A,B,C,D,R) - H(A,B,C) \ge 0$

\noindent with a factor of $861$: \\ $
H(A,B,C,D,T,U) + H(A,B,C,R,T,U) - H(A,B,C,D,R,T,U) - H(A,B,C,T,U) \ge 0$
 
\noindent with a factor of $522$: \\ $ 
H(A,B,C,D,S,U) + H(A,B,C,R,S,U) - H(A,B,C,D,R,S,U) - H(A,B,C,S,U) \ge 0$

\noindent with a factor of $191$: \\ $ 
H(A,D) + H(A,S) - H(A,D,S) - H(A) \ge 0$

\noindent with a factor of $191$: \\ $ 
H(A,C,D,U) + H(A,C,S,U) - H(A,C,D,S,U) - H(A,C,U) \ge 0$

\noindent with a factor of $104$: \\ $ 
H(A,C,D,T,U) + H(A,C,S,T,U) - H(A,C,D,S,T,U) - H(A,C,T,U) \ge 0$

\noindent with a factor of $778$: \\ $ 
H(A,C,D,R) + H(A,C,R,S) - H(A,C,D,R,S) - H(A,C,R) \ge 0$

\noindent with a factor of $3096$: \\ $ 
H(A,B,C,D) + H(A,B,C,S) - H(A,B,C,D,S) - H(A,B,C) \ge 0$

\noindent with a factor of $1337$: \\ $ 
H(A,B,C,D,R,T,U) + H(A,B,C,R,S,T,U) - H(A,B,C,D,R,S,T,U) - H(A,B,C,R,T,U) \ge 0$

\noindent with a factor of $474$: \\ $ 
H(B,D) + H(B,T) - H(B,D,T) - H(B) \ge 0$

\noindent with a factor of $237$: \\ $ 
H(B,C,D,U) + H(B,C,T,U) - H(B,C,D,T,U) - H(B,C,U) \ge 0$

\noindent with a factor of $711$: \\ $ 
H(A,D) + H(A,T) - H(A,D,T) - H(A) \ge 0$

\noindent with a factor of $46$: \\ $ 
H(A,B,D,U) + H(A,B,T,U) - H(A,B,D,T,U) - H(A,B,U) \ge 0$

\noindent with a factor of $341$: \\ $ 
H(A,B,D,R) + H(A,B,R,T) - H(A,B,D,R,T) - H(A,B,R) \ge 0$

\noindent with a factor of $104$: \\ $ 
H(A,B,C,D) + H(A,B,C,T) - H(A,B,C,D,T) - H(A,B,C) \ge 0$

\noindent with a factor of $873$: \\ $ 
H(A,B,C,D,S) + H(A,B,C,S,T) - H(A,B,C,D,S,T) - H(A,B,C,S) \ge 0$

\noindent with a factor of $769$: \\ $ 
H(A,B,C,D,R,U) + H(A,B,C,R,T,U) - H(A,B,C,D,R,T,U) - H(A,B,C,R,U) \ge 0$

\newpage

\noindent with a factor of $474$: \\ $ 
H(D,T) + H(T,U) - H(D,T,U) - H(T) \ge 0$

\noindent with a factor of $237$: \\ $ 
H(B,D) + H(B,U) - H(B,D,U) - H(B) \ge 0$

\noindent with a factor of $46$: \\ $ 
H(A,D,T) + H(A,T,U) - H(A,D,T,U) - H(A,T) \ge 0$

\noindent with a factor of $191$: \\ $ 
H(A,D,S,T) + H(A,S,T,U) - H(A,D,S,T,U) - H(A,S,T) \ge 0$

\noindent with a factor of $370$: \\ $
H(A,C,D,T) + H(A,C,T,U) - H(A,C,D,T,U) - H(A,C,T) \ge 0$
 
\noindent with a factor of $815$: \\ $ 
H(A,B,C,D) + H(A,B,C,U) - H(A,B,C,D,U) - H(A,B,C) \ge 0$

\noindent with a factor of $1722$: \\ $ 
H(A,B,C,D,S,T) + H(A,B,C,S,T,U) - H(A,B,C,D,S,T,U) - H(A,B,C,S,T) \ge 0$

\noindent with a factor of $341$: \\ $ 
H(R,U) + H(T,U) - H(R,T,U) - H(U) \ge 0$

\noindent with a factor of $607$: \\ $ 
H(C,R) + H(C,T) - H(C,R,T) - H(C) \ge 0$

\noindent with a factor of $208$: \\ $ 
H(B,C,D,R,S) + H(B,C,D,S,T) - H(B,C,D,R,S,T) - H(B,C,D,S) \ge 0$

\noindent with a factor of $195$: \\ $ 
B,C,D,R,S,U) + H(B,C,D,S,T,U) - H(B,C,D,R,S,T,U) - H(B,C,D,S,U) \ge 0$

\noindent with a factor of $948$: \\ $ 
H(A,R) + H(A,T) - H(A,R,T) - H(A) \ge 0$

\noindent with a factor of $963$: \\ $ 
H(A,C,R,S) + H(A,C,S,T) - H(A,C,R,S,T) - H(A,C,S) \ge 0$

\noindent with a factor of $293$: \\ $ 
H(A,B,C,R,U) + H(A,B,C,T,U) - H(A,B,C,R,T,U) - H(A,B,C,U) \ge 0$

\noindent with a factor of $403$: \\ $ 
H(C,D,R) + H(C,D,U) - H(C,D,R,U) - H(C,D) \ge 0$

\noindent with a factor of $476$: \\ $ 
H(A,C,R,S) + H(A,C,S,U) - H(A,C,R,S,U) - H(A,C,S) \ge 0$

\noindent with a factor of $39$: \\ $ 
H(A,B,D,R) + H(A,B,D,U) - H(A,B,D,R,U) - H(A,B,D) \ge 0$

\noindent with a factor of $849$: \\ $ 
A,B,D,R,S,T) + H(A,B,D,S,T,U) - H(A,B,D,R,S,T,U) - H(A,B,D,S,T) \ge 0$

\noindent with a factor of $948$: \\ $ 
H(R,S) + H(R,T) - H(R,S,T) - H(R) \ge 0$

\noindent with a factor of $208$: \\ $ 
H(C,D,S) + H(C,D,T) - H(C,D,S,T) - H(C,D) \ge 0$

\noindent with a factor of $341$: \\ $ 
H(B,C,D,R,S) + H(B,C,D,R,T) - H(B,C,D,R,S,T) - H(B,C,D,R) \ge 0$

\noindent with a factor of $191$: \\ $ 
H(A,D,S) + H(A,D,T) - H(A,D,S,T) - H(A,D) \ge 0$

\noindent with a factor of $607$: \\ $ 
H(A,C,R,S) + H(A,C,R,T) - H(A,C,R,S,T) - H(A,C,R) \ge 0$

\noindent with a factor of $208$: \\ $ 
H(A,C,D,S) + H(A,C,D,T) - H(A,C,D,S,T) - H(A,C,D) \ge 0$

\noindent with a factor of $522$: \\ $ 
H(A,B,C,S,U) + H(A,B,C,T,U) - H(A,B,C,S,T,U) - H(A,B,C,U) \ge 0$

\noindent with a factor of $293$: \\ $ 
H(A,B,C,D,R,S,U) + H(A,B,C,D,R,T,U) - H(A,B,C,D,R,S,T,U) - H(A,B,C,D,R,U) \ge 0$

\noindent with a factor of $208$: \\ $ 
H(C,D,S,T) + H(C,D,T,U) - H(C,D,S,T,U) - H(C,D,T) \ge 0$

\noindent with a factor of $403$: \\ $ 
H(B,C,D,R,S) + H(B,C,D,R,U) - H(B,C,D,R,S,U) - H(B,C,D,R) \ge 0$

\noindent with a factor of $136$: \\ $ 
H(A,B,D,S) + H(A,B,D,U) - H(A,B,D,S,U) - H(A,B,D) \ge 0$

\bigskip

\small

\noindent
The sum of these equalities  and inequalities (with the specified factors) is equal to  
\[ 
 4503 [ I(A;B|C) +  I(A;B|D) + I(C;D) - I(A;B) ] + 711 H(A,B,C,D) \ge 0,
\]
which is equivalent to 
\[ 
  I(A;B|C) +  I(A;B|D) + I(C;D) - I(A;B)  + (3/19) H(A,B,C,D) \ge 0.
\]
Therefore,
\[ 
  \frac{   I(A;B|C) +  I(A;B|D) + I(C;D) - I(A;B)  }{H(A,B,C,D)}  \ge -  {3}/{19}.
\]

\end{document}